\documentclass[runningheads]{llncs}
\usepackage{graphicx}
\usepackage{cleveref}
\usepackage{amsmath}
\usepackage{bookmark}
\usepackage{times}
\usepackage{xcolor}
\usepackage{verbatim}
\usepackage[utf8]{inputenc}
\usepackage{hyperref}
\usepackage{amssymb}
\usepackage{csquotes}
\usepackage{listings}
\usepackage{nameref}
\usepackage{svg}
\usepackage{calc}
\usepackage{newfloat}
\usepackage{makecell}
\usepackage{subfig}
\usepackage{multirow}
\usepackage{color}
\usepackage{float}
\restylefloat{table}
\definecolor{lightgray}{rgb}{.9,.9,.9}
\definecolor{darkgray}{rgb}{.4,.4,.4}
\definecolor{purple}{rgb}{0.65, 0.12, 0.82}

\lstdefinelanguage{JavaScript}{
  keywords={typeof, new, true, false, catch, function, return, null, catch, switch, var, if, in, while, do, else, case, break},
  keywordstyle=\color{blue}\bfseries,
  ndkeywords={class, export, boolean, throw, implements, import, this},
  ndkeywordstyle=\color{darkgray}\bfseries,
  identifierstyle=\color{black},
  sensitive=false,
  comment=[l]{//},
  morecomment=[s]{/*}{*/},
  commentstyle=\color{purple}\ttfamily,
  stringstyle=\color{red}\ttfamily,
  morestring=[b]',
  morestring=[b]"
}

\lstdefinestyle{js}{
  language=[5.1]JavaScript,
  basicstyle=\ttfamily,
  keywordstyle=\color{magenta},
  stringstyle=\color{blue},
  commentstyle=\color{black!50}
}

\begin{document}
\title{JS-son - A Lean, Extensible JavaScript Agent Programming Library}
%
%
%
\author{Timotheus Kampik\orcidID{0000-0002-6458-2252} \and \\
Juan Carlos Nieves\orcidID{0000-0003-4072-8795}}
     \institute{Umeå University,\\ 901 87 Umeå \\ Sweden \\
     \email{\{tkampik,jcnieves\}@cs.umu.se}}
\authorrunning{T. Kampik and J.C. Nieves}
%
%
\maketitle
\begin{abstract}
A multitude of agent-oriented software engineering frameworks exist, most of which are developed by the academic multi-agent systems community.
However, these frameworks often impose programming paradigms on their users that are challenging to learn for engineers who are used to modern high-level programming languages such as JavaScript and Python.
To show how the adoption of agent-oriented programming by the software engineering mainstream can be facilitated, we provide a lean JavaScript library prototype for implementing 
reasoning-loop agents. The library focuses on core agent programming concepts and refrains from imposing further restrictions on the programming approach.
To illustrate its usefulness, we show how the library can be applied to multi-agent systems simulations on the web, deployed to cloud-hosted function-as-a-service environments, and embedded in Python-based data science tools.

\end{abstract}
\section{Introduction}
Many multi-agent system (MAS) platforms have been developed by the scientific community~\cite{kravari2015survey}.
However, these platforms are rarely applied outside of academia, likely because they require the adoption of design paradigms that are fundamentally different from industry practices and do not integrate well with modern software engineering tool chains.
A recent expert report on the status quo and future of \emph{engineering multi-agent systems}\footnote{The report was assembled as a result of the EMAS 2018 workshop.} concludes that \textquote{many frameworks that are frequently used by the MAS community--for example Jason and JaCaMo--have not widely been adopted in practice and are dependent on technologies that are losing traction in the industry}~\cite{engineering-gsi-article-2019}.
Another comprehensive assessment of the current state of agent-oriented software engineering and its implications on future research directions is provided in Logan's \emph{Agent Programming Manifesto}~\cite{doi:10.1504/IJAOSE.2018.094374}.
Both the \emph{Manifesto} and the EMAS report recommend developing agent programming languages that are easier to use (as one of several ways to facilitate the impact of multi-agent systems research). \newpage
The EMAS report highlights, \emph{in particular}, the following issues:
\begin{enumerate}
     \item The tooling of academic agent programming lacks maturity for industry adoption. In particular, Logan states that \textquote{there is little incentive for developers to switch to current agent programming languages, as the behaviours that can be easily programmed are sufficiently simple to be implementable in mainstream languages with only a small overhead in coding time}\cite{doi:10.1504/IJAOSE.2018.094374}.
     \item Recent trends towards higher-level programming languages have found little consideration by the multi-agent systems community. In contrast, the machine learning community has embraced these programming languages, for example by providing frameworks like Tensorflow.js for JavaScript~\cite{smilkov2019tensorflow} and Keras for Python~\cite{Chollet:2017:DLP:3203489}.
     \item Consequently, agent programming lacks strong industry success stories.
\end{enumerate}
Based on these challenges, the following research directions can be derived:
\begin{enumerate}
     \item Provide agent programming tools that offer useful abstractions in the context of modern technology ecosystems/software stacks, without imposing unnecessarily complex design abstractions or niche languages onto developers.
     \item Embrace emerging technology ecosystems that are increasingly adopted by the industry, like Python for data science/machine learning and JavaScript for the web.
     \item Evaluate agent programming tools in the context of industry software engineering.
\end{enumerate}
While this work cannot immediately provide practical agent programming success stories, it attempts to provide a contribution to the development of tools and frameworks that are conceptually pragmatic in that they limit the design concepts and technological peculiarities they impose on their users and allow for a better integration into modern software engineering ecosystems.
We follow a pragmatic and lean approach: instead of creating a comprehensive multi-agent systems framework, we create \emph{JS-son}, a light-weight library that can be applied in the context of existing industry technology stacks and tool chains and requires little additional, MAS-specific knowledge.

The rest of this chapter is organized as follows.
The design approach for JS-son is described in Section~\ref{design}.
The architecture of JS-son, as well as the supported reasoning loops, are explained in Section~\ref{architecture}.
Subsequently, Section~\ref{programming} explains how to program JS-son agents using a small, step-by-step example.
Section~\ref{use-cases} elaborates on scenarios, in which using JS-son can be potentially beneficial; for some of the use case types, simple proof-of-concept examples are presented in Section~\ref{examples}.
Then, JS-son is put into the context of related work on agent programming libraries and frameworks in high-level programming languages in Section~\ref{related}.
Finally, Section~\ref{discussion} concludes the chapter by discussing limitations and future work.

\section{Design Approach}
\label{design}
Programming languages like Lisp and Haskell are rarely used in practice but have influenced the adoption of (functional) features in mainstream languages like JavaScript and C\#.
It is not uncommon that an intermediate adoption step is enabled by external libraries.
For example, before JavaScript's \texttt{array.prototype.includes} function was adopted as part of the ECMA Script standard\footnote{\url{https://www.ecma-international.org/ecma-262/7.0/\#sec-array.prototype.includes}}, a similar function (\texttt{contains} and its aliases \texttt{include} / \texttt{includes}) could already be imported with the external library \texttt{underscore}\footnote{\url{https://underscorejs.org/\#contains}}.
Analogously, JS-son takes the belief-desire-intention (BDI)~\cite{rao91a} architecture as popularized in the MAS community by frameworks like Jason~\cite{Bordini:2007:PMS:1197104} (as the name \emph{JS-son} reflects) and provides an abstraction of the BDI architecture (as well as support for other reasoning loops) as a \emph{plug and play} dependency for a widely adopted programming language.
Table~\ref{comparison} provides a side-by-side overview of the influence of the functional programming paradigm via Lisp's \texttt{MEMBER} function on JavaScript's \texttt{includes} function as an analogy to the influence of Jason's \texttt{(event, context, body)}-plans on JS-son's \texttt{(intention-condition, body)}-plans.
%
\begin{table}
     \centering
    \caption{Evolution of a Functional Feature from Lisp to JavaScript and Development of an Agent-oriented Feature from Jason to JS-son.}
    \label{comparison}
    \bgroup
     \def\arraystretch{1}
     \setlength\tabcolsep{10pt}
    \begin{tabular}{l|l|l}
                   & Functional Programming & Agent-oriented Programming \\
     \hline
     Source technology &  Lisp & Jason  \\
     Source feature,  & \texttt{MEMBER} function (list) & \makecell{(\texttt{event, context,} \\ \texttt{body)} plans}  \\
     Target technology & \multicolumn{2}{l}{\phantom{eeeeeeeeeeeeeeeeeeeee}JavaScript}  \\
     Target feature & \texttt{includes} functor (array) & \makecell{\texttt{(intention-condition,} \\ \texttt{body)} plans}  \\
     Library/extension & Lodash (\texttt{\_}) & JS-son  \\
     Standard feature & \texttt{includes} (ES2016) & none  \\
     \end{tabular}
     \egroup
\end{table}
To further guide the design and development of JS-son, we introduce three design principles that are--in their structure, as well as in their intend to avoid unnecessary overhead on the software (agent) engineering process--influenced by the \emph{Agile Manifesto}\footnote{\url{http://agilemanifesto.org/}}.
\begin{description}
     \item[Usability over intellectual elegance.] 
     JS-son provides a core framework for defining agents and their reasoning loops and environments, while allowing users to stick to pure JavaScript syntax and to apply their preferred libraries and design patterns to implement agent-agnostic functionality.
     \item[Flexibility over rigor.] Instead of proposing a \emph{one-size-fit-all} reasoning loop, JS-son offers flexibility in that it supports different approaches and is intended to remain open to evolve its reasoning loop as it matures.
     \item[Extensibility over out-of-the-box power.] To maintain JS-son as a concise library that can be adapted to a large variety of use cases while requiring little additional learning effort, we keep the JS-son core small and abstain from adding complex, special-purpose features, in particular if doing so imposed additional learning effort for JS-son users or required the use of third-party dependencies; \emph{i.e.}, we maintain a lean JS-son core module that is written in \emph{vanilla JavaScript} (does not require dependencies). Additional functionality can be provided as modules that extend the core and are managed as separate packages.
\end{description}

\section{Architecture and Reasoning Loops}
\label{architecture}
The library provides object types for creating agent and environment objects, as well as functions
for generating agent beliefs, desires, intentions, and plans~\footnote{The library--including detailed documentation, examples, and tests--is available at \url{https://github.com/TimKam/JS-son}.}. 

\begin{figure}
     \centering
     \subfloat[JS-son reasoning loop. The \emph{XOR} gateways allow for different reasoning loop approaches. The red sequence flows indicate the path of the belief-desire-intention-plan reasoning loop.]{\includegraphics[width=0.45\textwidth]{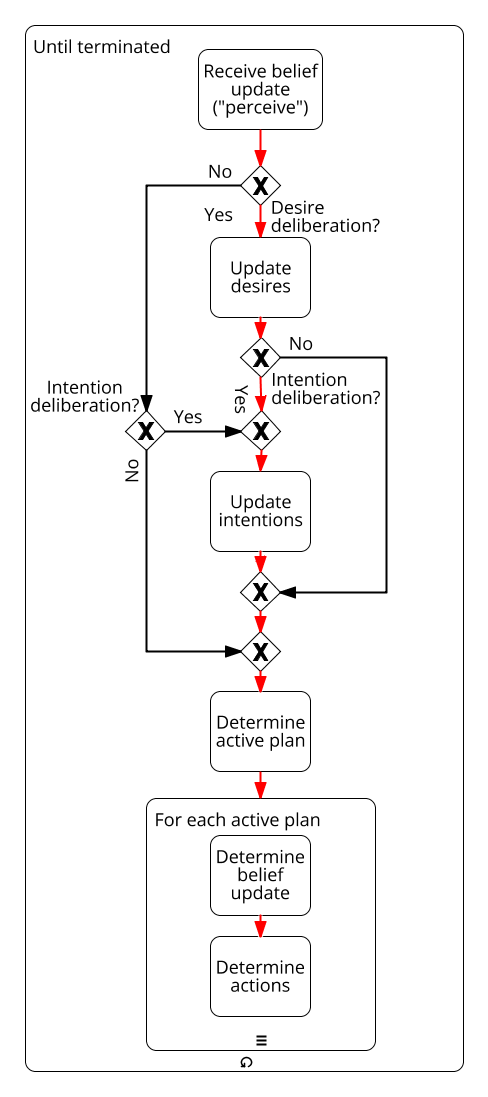}\label{deliberation_process}}
     \hspace{10pt}
     \centering
     \subfloat[JS-son environment: agent and state management process. The \emph{XOR} gateway allows for partially and fully observable environments.]{\includegraphics[width=0.45\textwidth]{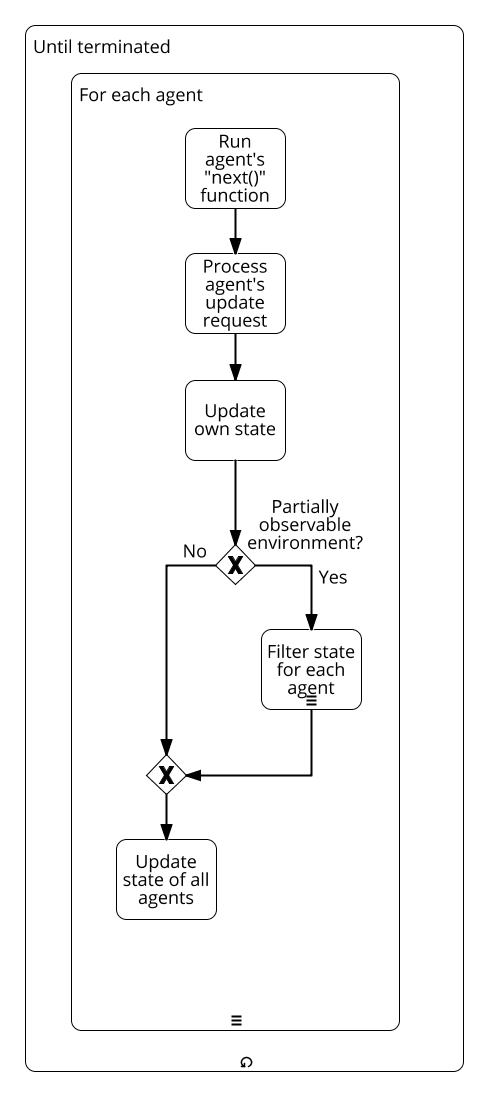}\label{environment_process}}
     \caption{JS-son reasoning and environment loop.}
 \end{figure}

\noindent The \textbf{agent} implements the BDI concepts as follows:
\begin{description}
     \item[Beliefs:] A belief can be any JavaScript Object Notation (JSON\footnote{\url{http://www.ecma-international.org/publications/files/ECMA-ST/ECMA-404.pdf}}) object or JSON data type (string, number, array, boolean, or $null$).
     \item[Desires:] Desires are generated dynamically by agent-specific desire functions that have a desire identifier assigned to them and determine the value of the desire based on the agent's current beliefs.
     \item[Intentions:] A \texttt{preference} function filters desires and returns \emph{intentions} - an array of JSON objects.
     \item[Plans:] A plan's \emph{head} specifies which intention needs to be active for the plan to be pursued. The plan body specifies how the plan should update the agent's beliefs and determines the actions the agent should issue to the environment.
\end{description}
Each agent has a \texttt{next()} function to run the following process:
\begin{enumerate}
     \item It applies the belief update as provided by the environment (see below).
     \item It applies the agent's preference function that dynamically updates the intentions based on the new beliefs; \emph{i.e.}, the agent is \emph{open-minded} (see Rao and Georgeff~\cite{rao91a}).
     \item It runs the plans that are active according to the updated intentions, while also updating the agent beliefs (if specified in the plans).
     \item It issues action requests that result from the plans to the environment.
\end{enumerate}
It is also possible to implement simpler belief-plan agents; \emph{i.e.}, as a plan's head, one can define a function that determines--based on the agent's current beliefs--if a plan should be executed. 
Alternatively, belief-desire-plan/belief-intention-plan reasoning loops are supported; these approaches bear similarity to the belief-goal-plan approach of the \emph{GOAL} language~\cite{hindriks2009programming}.
Figure~\ref{deliberation_process} depicts the reasoning loops that are supported by standard JS-son agents.

\noindent The \textbf{environment} contains the agents, as well as a definition of its own state. It executes the following instructions in a loop:
\begin{enumerate}
     \item It runs each agent's \texttt{next()} function.
     \item Once the agent's action request has been received, the environment processes the request. To determine which update requests should, in fact, be applied to the environment state, the environment runs the request through a filter function.
     \item When an agent's actions are processed, the environment updates its own state and the beliefs of all agents accordingly. Another filter function determines how a specific agent should \textquote{perceive} the environment's state.
\end{enumerate}
Figure~\ref{environment_process} depicts the environment's agent and state management process\footnote{In its current version, JS-son executes all steps \emph{synchronously}. Supporting the asynchronous execution, in particular of agent plans is future work, as discussed in Section~\ref{discussion}.}.
 %

\section{Implementing JS-son Agents}
\label{programming}
This section explains how to implement JS-son agents, by first giving a detailed explanation of the most important parts of the JS-son core API and then providing a programming tutorial.

\subsection{JS-son Core API}
The JS-son core API provides two major abstractions: one for agents and one for environments\footnote{Here, we only explain the core functionality for instantiating agents and environments. A comprehensive, continuously updated documentation of the JS-son API is available at \url{https://js-son.readthedocs.io/en/latest/}.}.
In addition, the agent requires the instantiation of beliefs, desires, and plans.
Note that intentions are generated dynamically, as is explained below.

\subsubsection*{Agents}
An agent is instantiated by calling the $Agent$ function with parameters that specify the agent's identifier (a text string), as well as its initial beliefs, desires, plans, and a \emph{preference function generator}.
Beliefs, desires, and plans are generated by the $Belief$, $Desire$, and $Plan$ functions, respectively.
Beliefs and desires consist of an identifier (key) and a body (value).
A belief body can be any valid JSON object or property (number, string, $null$, boolean, or array).
A desire body is a function that processes the agent's current beliefs and returns the processing result.
A plan has two functions; one as its \emph{body} and one as its \emph{head}.
The head determines--based on an agent's beliefs--if the plan body should be executed.
The body determines agent actions, as well as belief updates, taking the agent's beliefs as an optional input.
\emph{Intentions} are created by a preference function generator, a higher-order function that, based on the agents' current desires and beliefs, generates a function that reduces the agents' desires to intentions.
Table~\ref{agent} documents the $Agent$ function signature, whereas Tables~\ref{belief},~\ref{desire}, and~\ref{plan} document the signatures for the $Belief$, $Desire$, and $Plan$ functions, respectively.
\begin{table}
     \def\arraystretch{1.5}
     \setlength\tabcolsep{10pt}
     \caption{Function signature of the JS-son $Agent$ and its components.}
     \label{agent-all}
     \centering
\subfloat[JS-son $Agent$ function signature]{
     \centering
     \begin{tabular}{ l | l | l }
          \hline
          \textbf{Name} & \textbf{Type} & \textbf{Description} \\ \hline
          $id$ & String & Unique identifier of the agent \\
          $beliefs$ & Object & Initial beliefs of the agents \\
          $desires$ & Object & The agent's desires \\
          $plans$ & Array & The agent's plans \\ \hline
          $preferenceFunctionGenerator$ & Array & \makecell{Preference function generator; by default \\ (if no function is provided), the preference \\ function turns all desires into intentions} \\ \hline
          \textbf{Returns} & Object & JS-son Agent object \\
          \hline
        \end{tabular}\label{agent}
} \\
\subfloat[JS-son $Belief$ function signature]{
     \centering
     \begin{tabular}{ l | l | l }
      \hline
      \textbf{Name} & \textbf{Type}  & \textbf{Description} \\ \hline
      $id$ & String & Unique identifier of the belief \\
      $value$ &  \makecell{Any \\ (needs to be valid JSON object or JSON value)} & The belief's initial value \\ \hline
      \textbf{Returns} & Object & JS-son Belief object \\
      \hline
    \end{tabular}\label{belief}
} \\
\subfloat[JS-son $Desire$ function signature]{
     \centering
     \begin{tabular}{ l | l | l }
          \hline
          \textbf{Name} & \textbf{Type}  & \textbf{Description} \\ \hline
          $id$ & String & Unique identifier of the belief \\
          $body$ & Function & Function for computing the desires value based on current beliefs \phantom{ee}\\ \hline
          \textbf{Returns} & Object & JS-son Desire object \\
          \hline
        \end{tabular}\label{desire}
} \\
\subfloat[JS-son $Plan$ function signature]{
     \centering
     \begin{tabular}{ l | l | l }
          \hline
          \textbf{Name} & \textbf{Type}  & \textbf{Description} \\ \hline
          $head$ & Function & Determines if plan is active \\
          $body$ & Function & Determines the execution of actions and update of beliefs \phantom{eeeeeeeee} \\ \hline
          \textbf{Returns} & object & Plan object \\
          \hline
     \end{tabular}\label{plan}
}
\end{table}

\subsubsection*{Environment}
The environment is generated by the $Environment$ function that takes as its input an array of JS-son agents, an initial state definition (JSON object), and functions for updating the environment's state, visualizing it, and pre-processing (filtering or manipulating) it before exposing the state to the agents.
The $update$ function processes the agents' actions; for each agent, it determines how the environment's state should be updated, based on the current state, the agent's actions, and the agent's identifier.
The state update is then visualized as specified by the $render$ function. In case a visualization is not necessary, the default $render$ function makes the environment log each iteration's state to the console.
The $stateFilter$ function filters or manipulates the state as perceived by a particular agent, based on this agent's identifier and its current beliefs; by default (if no $stateFilter$ function is specified), the state is returned unfiltered to the agent(s).
Table~\ref{env} documents the environment's function signature.
\begin{table}
     \def\arraystretch{1.5}
     \setlength\tabcolsep{10pt}
     \centering
    \caption{JS-son $Environment$ function signature}
    \label{env}
    \begin{tabular}{ l | l | l }
     \hline
     \textbf{Name} & \textbf{Type} & \textbf{Description} \\ \hline
     $agents$ & Array of JS-son agents & Agents that the environment is managing \\
     $state$ & Object & Initial state of the environment \\
     $update$ & Function & \makecell{Processes agent actions and updates the \\ environment's state} \\
     $render$ & Function & Visualizes the environment's current state \\
     $stateFilter$ & Function & \makecell{Filters/manipulates the state that agents \\ should perceive} \\\hline
     \textbf{Returns} & Object & Plan object \\
     \hline
   \end{tabular}
\end{table}

\subsection{Tutorial}
The tutorial explains how to program \emph{belief-plan} agents using a minimal example\footnote{Tutorials that present more complex examples are available in the JS-son project documentation \url{https://js-son.readthedocs.io}.}.
Running the example requires the creation of a new Node.js project (\texttt{npm init}), the installation of the \texttt{js-son-agent} dependency, and the import of the \texttt{JS-son} library.
\begin{lstlisting}
const {
     Belief,
     Plan,
     Agent,
     Environment } = require('js-son-agent')
\end{lstlisting}
The tutorial implements the \emph{Jason room example}\footnote{\url{https://github.com/jason-lang/jason/tree/master/examples/room}} with JS-son.
In the example, three agents are in a room:
\begin{enumerate}
     \item A porter that locks and unlocks the room's door if requested;
     \item A paranoid agent that prefers the door to be locked and asks the porter to lock the door if this is not the case;
     \item A claustrophobe agent that prefers the door to be unlocked and asks the porter to unlock the door if this is not the case.
\end{enumerate}

The simulation runs twenty iterations of the scenario. In an iteration, each agent acts once.
All agents start with the same beliefs. The belief with the ID \texttt{door} is assigned the object \texttt{\{locked: true\}}; \emph{i.e.}, the door is locked. Also, nobody has so far requested any change in door state (\texttt{requests: []}).
\begin{lstlisting}
const beliefs = {
  ...Belief('door', { locked: true }),
  ...Belief('requests', [])
}
\end{lstlisting}
Now, we define the porter agent. The porter has the following plans:
\begin{enumerate}
     \item If it does not believe the door is locked and it has received a request to lock the door (head), lock the door (body).
     \item If it believes the door is locked and it has received a request to unlock the door (head), unlock the door (body).
\end{enumerate}
\begin{lstlisting}
const plansPorter = [
     Plan(
          beliefs =>
               !beliefs.door.locked &&
               beliefs.requests.includes('lock'),
          () => [{ door: 'lock' }]
     ),
     Plan(
          beliefs =>
               beliefs.door.locked &&
               beliefs.requests.includes('unlock'),
          () => [{ door: 'unlock' }]
     )
]
\end{lstlisting}
We instantiate a new agent with the belief set and plans. Because we are not making use of desires in this simple belief-plan scenario, we pass an empty object as the agent's desires.
\begin{lstlisting}
const porter = new Agent('porter', beliefs, {}, plansPorter)
\end{lstlisting}
Next, we create the paranoid agent with the following plans:
\begin{enumerate}
     \item If it does not belief the door is locked (head), it requests the door to be locked (body).
     \item If it beliefs the door is locked (head), it broadcasts a thank you message for locking the door (body).
\end{enumerate}
\newpage
\begin{lstlisting}
const plansParanoid = [
  Plan(
    beliefs => !beliefs.door.locked,
    () => [{ request: 'lock' }]
  ),
  Plan(
    beliefs => beliefs.door.locked,
    () => [{ announce: 'Thanks for locking the door!' }]
  )
]

const paranoid = new Agent('paranoid', beliefs, {}, plansParanoid)  
\end{lstlisting}
The last agent we create is the paranoid one. It has these plans:
\begin{enumerate}
     \item If it beliefs the door the door is locked (head), it requests the door to be unlocked (body).
     \item If it does not belief the door is locked (head), it broadcasts a thank you message for unlocking the door (body).
\end{enumerate}
\begin{lstlisting}
const plansClaustrophobe = [
     Plan(
          beliefs => beliefs.door.locked,
          () => [{ request: 'unlock' }]
     ),
     Plan(
          beliefs => !beliefs.door.locked,
          () => [{ announce: 'Thanks for unlocking the door!' }]
     )
]
     
const claustrophobe = new Agent(
     'claustrophobe',
     beliefs,
     {},
     plansClaustrophobe
)
\end{lstlisting}
Now, as we have defined the agents, we need to specify the environment. First, we set the environments state, which is--in our case--consistent with the agents' beliefs.
\begin{lstlisting}
const state = {
  door: { locked: true },
  requests: []
}
\end{lstlisting}
To define how the environment processes agent actions, we implement the \texttt{updateState} function. The function takes an agent's actions, as well as the agent's identifier and the current state to determine the environment's state update that is merged into the new state \texttt{state = { ...state, ...stateUpdate }}.
\begin{lstlisting}
const updateState = (actions, agentId, currentState) => {
     const stateUpdate = {
          requests: currentState.requests
     }
     actions.forEach(action => {
          if (action.some(action => action.door === 'lock')) {
               stateUpdate.door = { locked: true }
               stateUpdate.requests = []
               console.log(`${agentId}: Lock door`)
          }
          if (action.some(action => action.door === 'unlock')) {
               stateUpdate.door = { locked: false }
               stateUpdate.requests = []
               console.log(`${agentId}: Unlock door`)
          }
          if (action.some(action => action.request === 'lock')) {
               stateUpdate.requests.push('lock')
               console.log(`${agentId}: Request: lock door`)
          }
          if (action.some(action => action.request === 'unlock')) {
               stateUpdate.requests.push('unlock')
               console.log(`${agentId}: Request: unlock door`)
          }
          if (action.some(action => action.announce)) {
               console.log(`${agentId}: ${
                    action.find(
                         action => action.announce
                    ).announce
               }`)
          }
     })
     return stateUpdate
}
\end{lstlisting}
To simulate a partially observable world, we can specify the environment's \texttt{stateFilter} function, which determines how the state update should be shared with the agents.
However, in our case we simply communicate the whole state update to all agents, which is also the default behavior of the environment, if no \texttt{stateFilter} function is specified.
\begin{lstlisting}
const stateFilter = state => state
\end{lstlisting}
We instantiate the environment with the specified agents, state, update function, and filter function.
\begin{lstlisting}
const environment = new Environment(
     [paranoid, claustrophobe, porter],
     state,
     updateState,
     stateFilter
)
\end{lstlisting}
Finally, we run 20 iterations of the scenario.
\begin{lstlisting}
environment.run(20)
\end{lstlisting}

\section{Potential Use Cases}
\label{use-cases}
We suggest that JS-son can be applied in the following use cases: 

\begin{description}
     \item[Data science.] With the increasing relevance of large-scale and semi-automated statistical analysis (\textquote{data science}) in industry and academia, a new set of technologies has emerged that focuses on pragmatic and flexible usage and treats traditional programming paradigms as second-class citizens. JS-son integrates well with Python- and Jupyter notebook\footnote{\url{https://jupyter.org/}.}-based data science tools, as shown in \hyperref[demo-1]{Demonstration 1}.
     \item[Web development.] Web front ends implement functionality of growing complexity; often, large parts of the application are implemented by (browser-based) clients. As shown in \hyperref[demo-2]{Demonstration 2}, JS-son allows embedding BDI agents in single-page web applications, using the tools and paradigms of web development.
     \item[Education.] Programming courses are increasingly relevant for educating students who lack a computer science background. Such courses are typically taught in high-level languages that enable students to write working code without knowing all underlying concepts. In this context, JS-son can be used as a tool for teaching MAS programming.
     \item[Internet-of-Things (IoT)] Frameworks like Node.js\footnote{\url{https://nodejs.org/}} enable the rapid development of IoT applications, as a large ecosystem of libraries leaves the application developer largely in the role of a system integrator. JS-son is available as a Node.js package.
     \item[Function-as-a-Service.] The term \emph{serverless}~\cite{baldini2017serverless} computing refers to information technology that allows application developers to deploy their code via the infrastructure and software ecosystem of third-party providers without needing to worry about the technical details of the execution environment. The provision of \emph{serverless} computing services is often referred to as \emph{Function-as-a-Service} (FaaS). Most FaaS providers, like Heroku\footnote{\url{https://devcenter.heroku.com/articles/getting-started-with-nodejs}}, Amazon Web Services Lamda\footnote{\url{https://docs.aws.amazon.com/lambda/latest/dg/nodejs-prog-model-handler.html}}, and Google Cloud Functions\footnote{\url{https://cloud.google.com/functions/docs/concepts/nodejs-8-runtime}}, provide Node.js support for their service offerings and allow for the deployment of JavaScript functions with little setup overhead. Consequently, JS-son can emerge as a convenient tool to develop agents and multi-agent systems that are then deployed as \emph{serverless} functions.
     For a running example, see Subsection~\ref{demo-4}.
\end{description}

\section{Examples}
\label{examples}
We provide four demonstrations that show how JS-son can be applied.
The code of all demonstration is available in the JS-son project repository (\url{https://github.com/TimKam/JS-son}).

\subsection{JS-son meets Jupyter}
\label{demo-1}
The first demonstration shows how JS-son can be integrated with data science tools, \emph{i.e.}, with Python libraries and Jupyter notebooks\footnote{The Jupyter notebook is available on GitHub at \url{http://s.cs.umu.se/lmfd69} and on a Jupyter notebook service platform at \url{http://s.cs.umu.se/girizr}.}. As a simple proof-of-concept example, we simulate opinion spread in an agent society and run an interactive data visualization.
The example simulates the spread of a single boolean belief among 100 agents in environments with different \emph{biases} regarding the facilitation of the different opinion values. Belief spread is simulated as follows:
\begin{enumerate}
     \item The scenario starts with each agent announcing their beliefs.
     \item In each iteration, the environment distributes two belief announcements to each agent. Based on these beliefs and possibly (depending on the agent type) the past announcements the agent was exposed to, each agent announces a new belief: either $true$ or $false$.
\end{enumerate}
The agents are of two different agent types (\emph{volatile} and \emph{introspective}):
\begin{description}
     \item[Volatile.] Volatile agents only consider their current belief and the latest belief set they received from the environment when deciding which belief to announce. Volatile agents are "louder", \emph{i.e.}, the environment is more likely to spread beliefs of volatile agents. We also add bias to the announcement spread function to favor true announcements.
     \item[Introspective.] In contrast to volatile agents, introspective agents consider the past five belief sets they have received, when deciding which belief they should announce. Introspective agents are "less loud", \emph{i.e.}, the environment is less likely to spread beliefs of volatile agents.
\end{description}
The agent type distribution is $50, 50$. However, 30 volatile and 20 introspective agents start with $true$ as their belief, whereas 20 volatile and 30 introspective agents start with $false$ as their belief.
Figure~\ref{jupyter} shows an excerpt of the Juypter notebook.

\subsection{JS-son in the Browser}\label{demo-2}
The second demonstration presents a JS-son port of \emph{Conway's Game of Life}.
It illustrates how JS-son can be used as part of a web frontend.
In this example, JS-son is fully integrated into a JavaScript build and compilation pipeline that allows writing modern, idiomatic JavaScript code based on the latest ECMAScript specification, as it compiles this code into cross-browser compatible, \emph{minified} JavaScript.
The demonstration makes use of JS-son's simplified belief-plan approach\footnote{The simulation is available at \url{http://s.cs.umu.se/chfbk2}.}.
Each Game of Life \emph{cell} is represented by an agent that has two beliefs: its own state (active or inactive) and the number of its \emph{active} neighbors.
At each simulation tick, the agent decides based on its beliefs, if it should register a change in its status (from active to inactive or vice versa) with the environment.
After all agents have registered their new status, the environment updates the global game state accordingly and passes the new number of active neighbors to each agent.
Figure~\ref{game_of_life} depicts the \emph{Game of Life} application.

\subsection{Learning JS-son Agents}\label{demo-3}
The third demonstration shows how \emph{learning} JS-son agents can be implemented in a browser-based grid world\footnote{This grid world is an adaptation of an environment in which learning JS-son agents are rewarded based on a specific, \emph{fair} game-theoretical equilibrium in a given state, as presented by Kampik and Spieker~\cite{kampikSpieker2019sais}.}.
The example instantiates agents in a $20 \times 20$ field grid world \emph{arena} with the following field types:
\begin{itemize}
     \item \emph{Mountain} fields that the agents cannot pass.
     \item \emph{Money} fields that provide a \emph{coin} to an agent that approaches them (the agent needs to move onto the field, but the environment will return a coin and leave the agent at its current position).
     \item \emph{Repair} fields that provide damaged agents with one additional health unit when approached (again, the agent needs to move onto the field, but the environment will return a health unit and leave the agent at its current position).
     \item \emph{Plain} fields that can be traversed by an agent if no other agent is present on the field. If another agent is already present, the environment will reject the move, but decrease both agents' health by 10. When an agent's health reaches (or goes below) zero, it is punished by a withdrawal of 100 coins from its stash. 
\end{itemize}
The agents are trained \emph{online} (no model is loaded/persisted) using deep Q-learning through an experimental JS-son learning extension.
Figure~\ref{gridworld} shows the agents in the grid world arena. 
%
%
\subsection{Serverless JS-son Agents}\label{demo-4}
\begin{figure}
     \centering
     \subfloat[Analysis of a JS-son multi-agent simulation in a Jupyter Notebook.]{\includegraphics[width=0.45\textwidth]{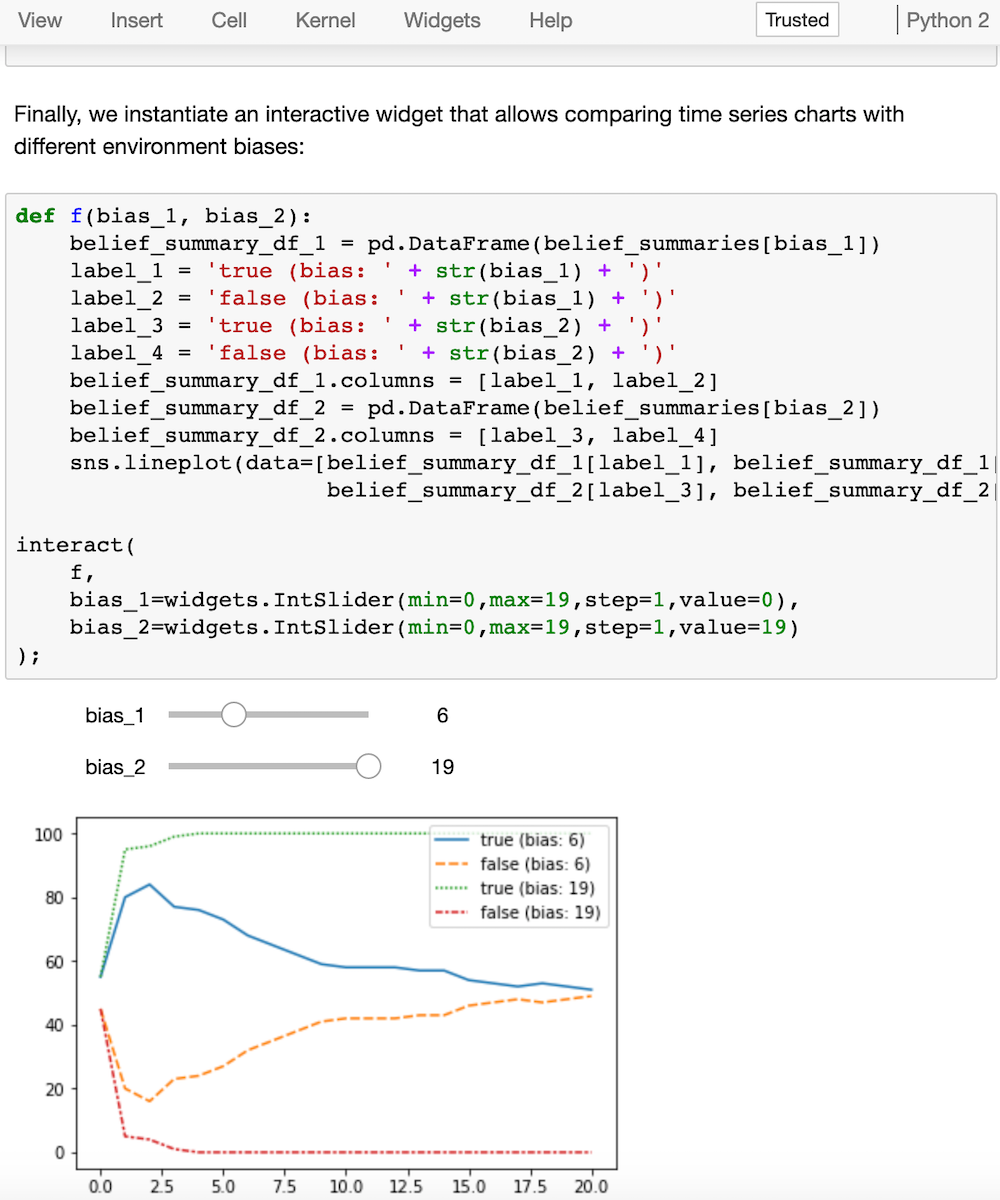}\label{jupyter}}
     \hspace{10pt}
     \centering
     \subfloat[JS-son: Conway's Game of Life, implemented as a web application. 
     ]{\includegraphics[width=0.45\textwidth]{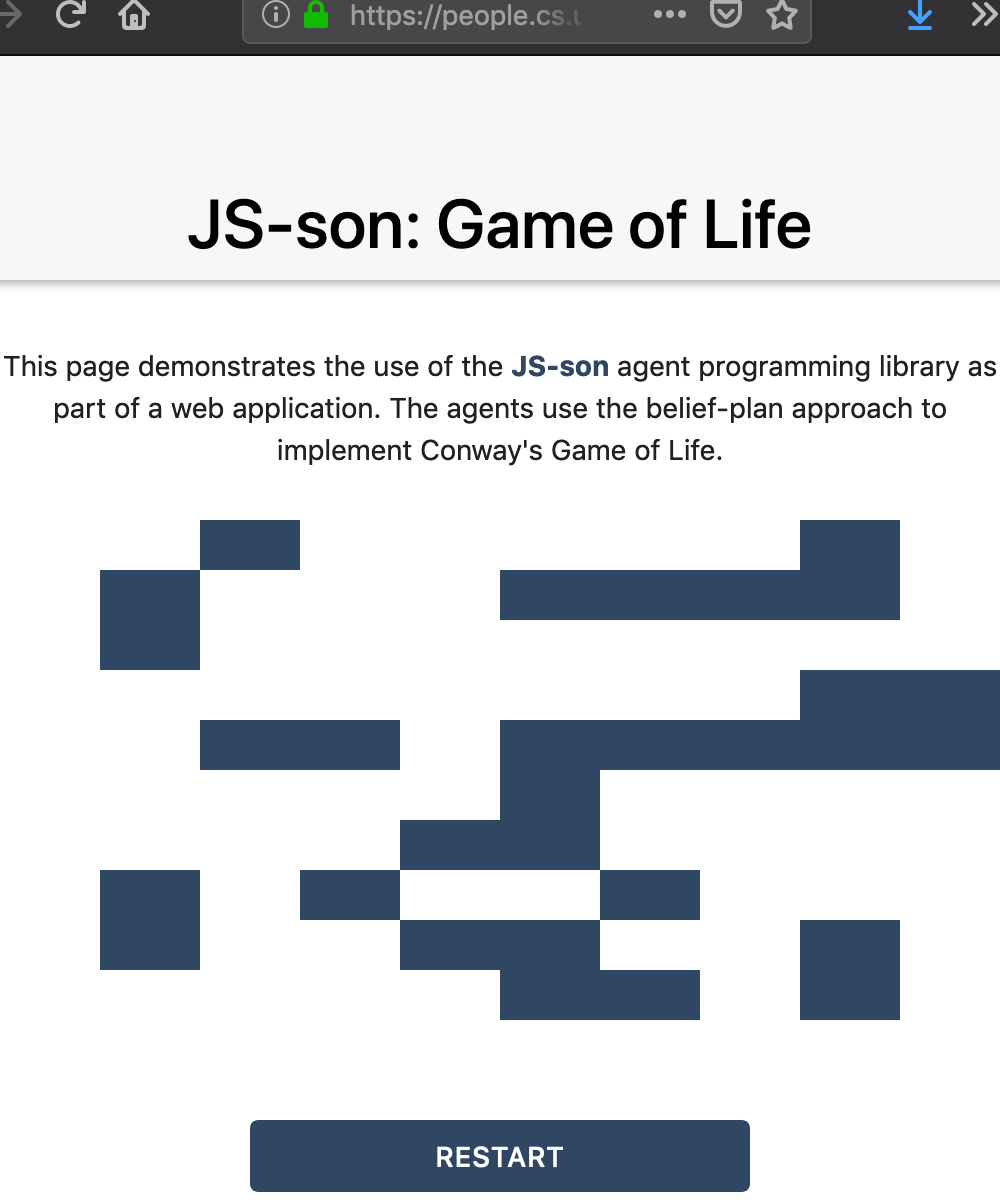}\label{game_of_life}}
     \hspace{10pt}
     \centering
     \subfloat[JS-son agents in a grid world.]{\includegraphics[width=0.45\textwidth]{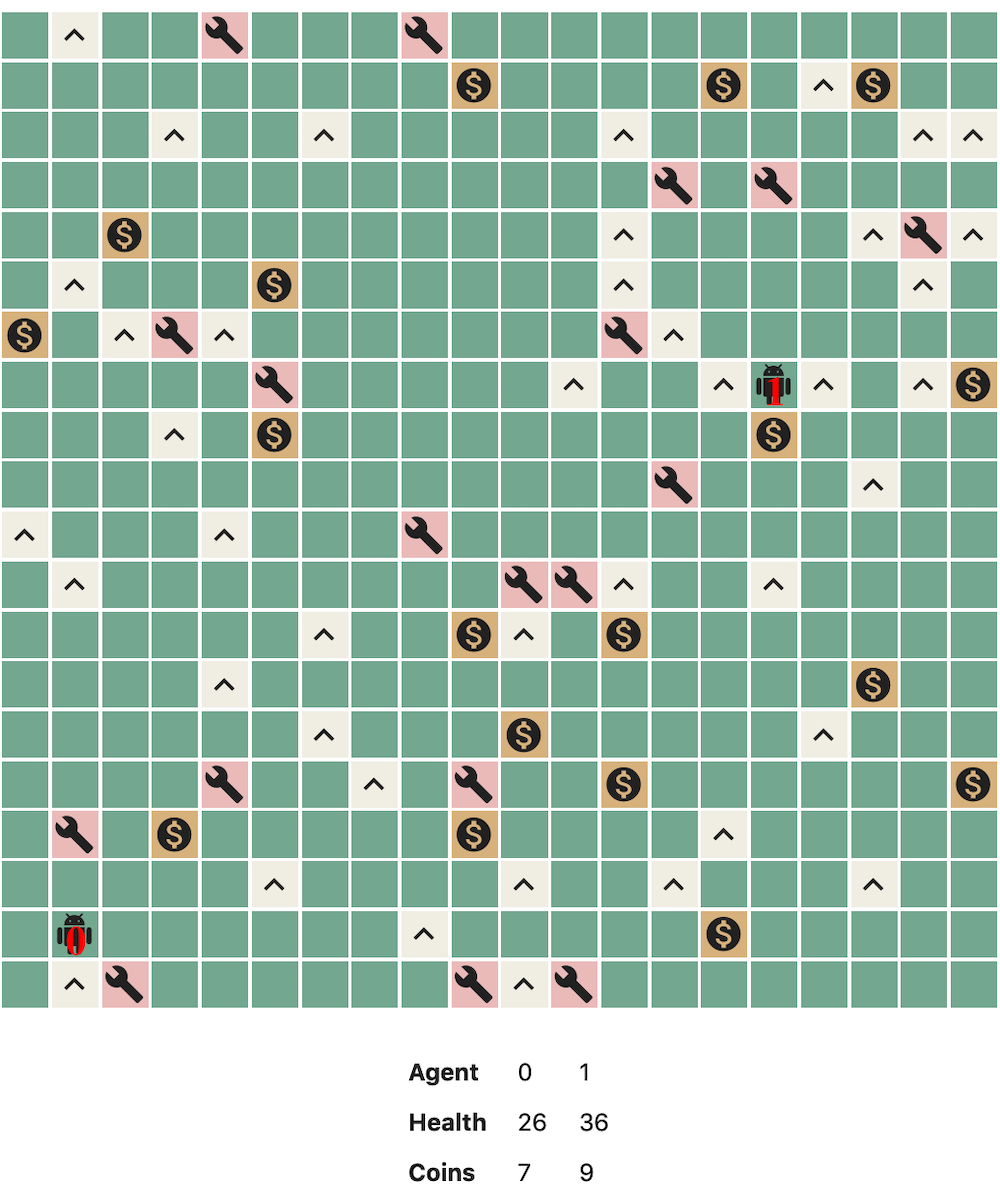}\label{gridworld}}
     \hspace{10pt}
     \centering
     \subfloat[JS-son multi-agent system, deployed as a Google Cloud Function.]{\includegraphics[width=0.45\textwidth]{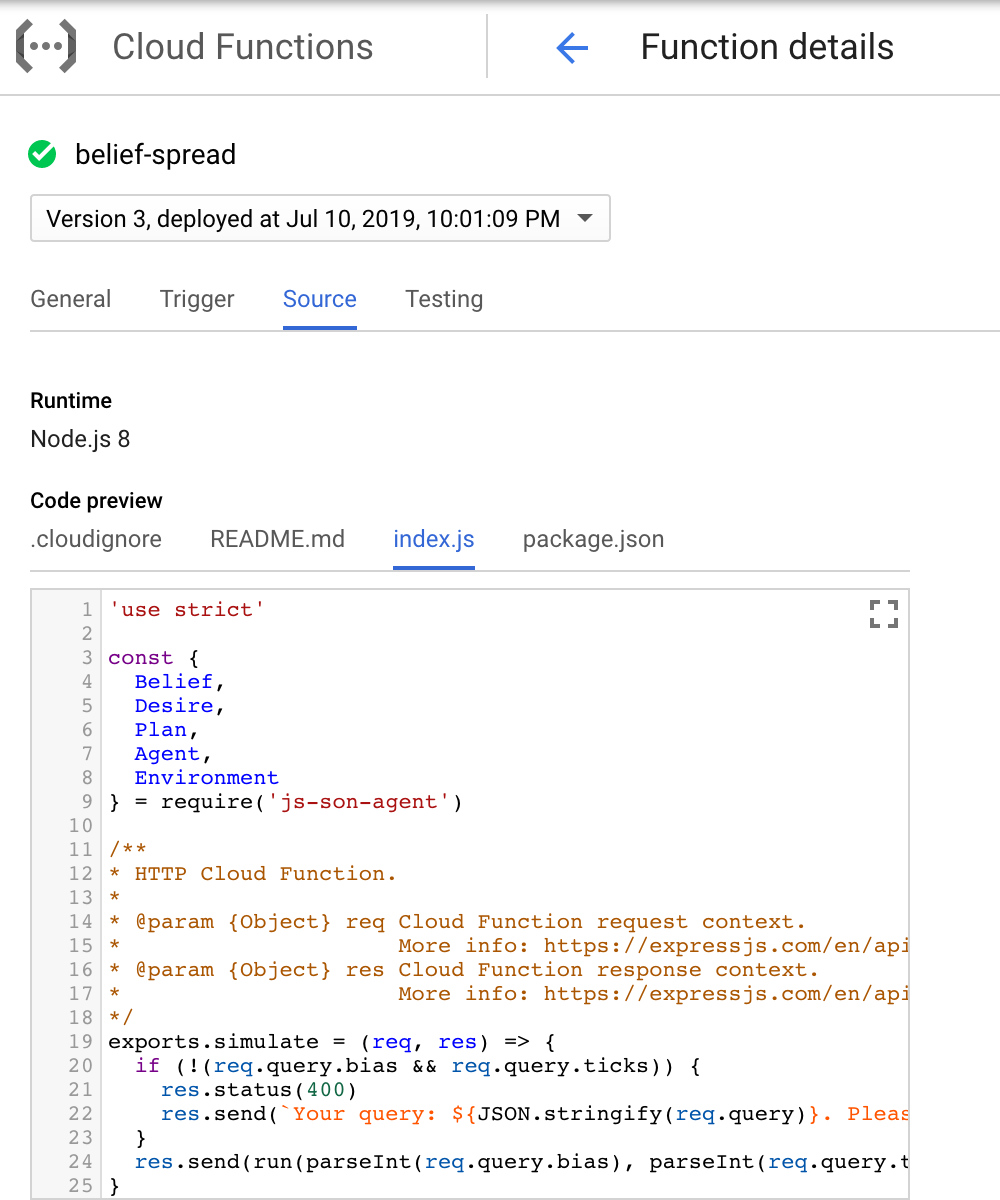}\label{cloud}}
     \caption{JS-son example applications.}
 \end{figure}

The fourth demonstration shows how JS-son agents can be deployed to \emph{Function-as-a-Service} providers.
It is based on the belief spread simulation as introduced in the first demonstration (see Subsection~\ref{demo-1}).
The multi-agent simulation is wrapped in a request handler and provided as a Node.js project that is configured to run as a \emph{Google Cloud Function}.
The request handler accepts HTTP(S) requests against the \texttt{simulate} endpoint. The request \emph{method} (\emph{e.g.}, \texttt{GET}, \texttt{POST}, \texttt{PUT}) is ignored by the handler.
Upon receiving the request, the handler runs the simulation for the specified number of \emph{ticks}, configuring the \emph{bias} in the agent society as specified by the corresponding request parameter (the higher the bias, the stronger the facilitation of \emph{true} announcements).
An example request against a fictional FaaS instance could be sent using the \texttt{curl} command line tool as specified in the code snippet below.
\begin{lstlisting}
curl -X GET 'https://instance.faas.net/simulation/simulate?ticks=20&bias=5'
\end{lstlisting}
Figure~\ref{cloud} depicts the simulation in the Google Cloud Functions management user interface.
%

\section{Related Work}
\label{related}
Over the past two decades, a multitude of agent-oriented software engineering frameworks emerged (see, \emph{e.g.}, Kravari and Bassiliades~\cite{kravari2015survey}).
However, most of these frameworks do not target higher-level programming languages like Python and JavaScript.
In this section, we provide a brief overview of three agent programing frameworks--\emph{osBrain}, \emph{JAM}, and \emph{Eve} that are indeed written in and for these two languages. We then highlight key differences to our library.

\subsection{osBrain}
osBrain\footnote{\url{https://osbrain.readthedocs.io/en/stable/about.html}} is a Python library for developing multi-agent systems.
Although osBrain is written in a different language than JS-son, it is still relevant for the comparison because it is \emph{i)} written in a higher level programming language of a similar generation and \emph{ii)} somewhat actively maintained\footnote{As of March 2020, the last update to the source of \emph{Eve} dates back more than 2.5 years to August 2017 (\url{https://github.com/enmasseio/evejs/}); the last update of the documentation of \emph{JAM}--whose source code is not available--dates back more than 1.5 years to August 2018 (\url{http://www.bsslab.de/?Software/jam}). In contrast the last update of the osBrain source and documentation dates back roughly one year to April 2019 (\url{https://github.com/opensistemas-hub/osbrain}).}.
Initially developed as an automated trading software backbone, the focus of osBrain lies on the provision of an agent-oriented communication framework. 
No framework for the agents internal reasoning loop is provided, \emph{i.e.} osBrain does not provide BDI support.
Also, osBrain dictates the use of a specific communication protocol and library, utilizing the message queue system \emph{ZeroMQ}~\cite{hintjens2013zeromq}.

\subsection{JavaScript Agent Machine (JAM)}
Bosse introduces the \emph{JavaScript Agent Machine} (JAM), which is a \textquote{mobile multi-agent system[...] for the Internet-of-Things and clouds}~\cite{bosse2016mobile}. \newpage
\noindent Some of JAM's main features and properties are, according to its documentation\footnote{\url{http://www.bsslab.de/assets/agents.html}}:
\begin{itemize}
     \item Performance: through third-party libraries, JAM agents can be compiled to Bytecode that allows for performant execution in low-resource environments;
     \item Mobility and support for heterogenous environments: agent instances can be moved between physical and virtual nodes at run-time; 
     \item Machine learning capabilities, through integration with a machine learning service platform; however, no details on how this service can be accessed are provided in the documentation.
\end{itemize}
In its initial version, JAM agents required the use of a JavaScript-like language that is syntactically not fully compliant with any standard JavaScript/ECMAScript version~\cite{bosse2015unified}.
However, in its latest version, it is possible to implement agent in syntactically valid JavaScript.
With its focus on agent orchestration, deployment, and communications, JAM's agent internals are based on \emph{activity-transition graphs}, which implies that its functionality overlaps little with JS-son.
Another point of distinction is that the JAM source code is not openly available; instead, the JAM website\footnote{\url{http://www.bsslab.de/?Software/jam}} provides a set of installers and libraries and software development kits for different platforms that can be used as black-box dependencies. 

\subsection{Eve}
De Jong \emph{et al.}~\cite{de2013eve} present \emph{Eve}, a multi-agent platform for agent discovery and communications.
It is available as both a Java and a JavaScript implementation.
Similar to osBrain, Eve's core functionality is an agent-oriented, unified abstraction on different communication protocols; it does not define agent internals like reasoning loops and consequently does not follow a belief-desire-intention approach.
Eve is provided as Node.js package\footnote{\url{https://www.npmjs.com/package/evejs}}, but as of March 2020, the installation fails and the Node Package Manager (npm) reports 11 known security vulnerabilities upon attempted installation.
Still, Eve is in regard to its technological basis similar to JS-son.
With its difference in focus--on agent discovery and communications in contrast to JS-son's reasoning loops--Eve could be, if maintenance issues will be addressed, a potential integration option that a JS-son extension can provide.

\subsection{Comparison - Unique JS-son Features}
To summarize the comparison, we list three unique features that distinguish JS-son from the aforementioned frameworks.
\begin{description}
     \item[Reasoning loop focus with belief-desire-intention support.]
     Of the three frameworks, only JAM provides a dedicated way to frame the reasoning loop of implemented agents, using activity-transition graphs.
     Still, the core focus of \emph{all three} libraries is on communication and orchestration, which contrasts the focus of JS-son as a library that has a reasoning loop framework at its core and aims to be largely agnostic to specific messaging and orchestration approaches.
     \item[Full integration with the modern JavaScript ecosystem.]
     As shown in Section~\ref{examples}, JS-son fully integrates with the JavaScript ecosystem across runtime environments.
     This is in particular a contrast to JAM, which 
     provides installers that obfuscate the proprietary source code and require a non-standard installation process.
     This can potentially hinder integration into existing software ecosystems that rely on \emph{industry standard} approaches to dependency management for continuous integration and delivery purposes.
     While Eve attempts to provide an integration that allows for a convenient deployment in different environments, for example through continuous integration pipelines, it does in fact not provide a working, stable, and secure installation package. 
     \item[Dependency-free and open source code.]
     JS-son is a light-weight, open source library that does not ship any dependencies in its core version, but rather provides modules that require dependencies as \emph{extensions}.
     In contrast, adopting JAM requires reliance on closed/obfuscated source code, whereas osBrain and Eve require a set of dependencies, which are in the case of Eve--as explained before--not properly managed.
\end{description}
\section{Conclusions and Future Work}
\label{discussion}
This chapter presents a lean, extensible library that provides simple abstractions for JavaScript-based agent programming, with a focus on reasoning loop specification.
To further increase the library's relevance for researchers, teachers, and practitioners alike, we propose the following work:
\begin{description}
     \item[Support a distributed environment and interfaces to other MAS frameworks.]
     It \\ makes sense to enable JS-son agents and environments to act in distributed systems and communicate with agents of other types, without requiring extensive customization by the library user. A possible way to achieve this is supporting the open standard agent communication language FIPA ACL\footnote{\url{http://www.fipa.org/specs/fipa00061/index.html}}. However, as highlighted in a previous publication~\cite{nieves2013intelligence}, FIPA ACL does not support communication approaches that have emerged as best practices for real-time distributed systems like \emph{publish-subscribe}.
     Also, the application of JS-son in a distributed context can benefit from the enhancement of agent-internal behavior, for example through a feature that supports the asynchronous execution of plans.
     \item[Implement a reasoning extension.]
     To facilitate JS-son's reasoning abilities, additional JS-son extensions can be developed. From an applied perspective, integrations with business rules engines can bridge the gap to traditional enterprise software, whereas a JS-son extension for \emph{formal argumentation} (see, \emph{e.g.}, Bench-Capon and Dunne~\cite{bench2007argumentation}) can be of value for the academic community.
     \item[Move towards real-world usage.] To demonstrate the feasibility of JS-son, it is important to apply the library in advanced scenarios. Considering the relatively small technical overhead JS-son agents imply, the entry hurdle for a development team to adopt JS-son is low, which can facilitate real-world adoption. Still, future work needs to evaluate how useful the abstractions JS-son provides are for industry software engineers.
     \item[Implement a Python port.] While JS-son can be integrated with the Python ecosystem, for example via Jupyter notebooks, doing so implies technical overhead and requires knowledge of two programming languages\footnote{Also, the module that allows for Node.js-Python interoperability (\url{https://github.com/pixiedust/pixiedust_node}) has some limitations, \emph{i.e.} it lacks Python 3 support.}. To facilitate the use of agents in a data science and machine learning context, we propose the implementation of \emph{Py\_son}, a Python port of JS-son.
\end{description}
\subsubsection{Acknowledgements}
The authors thank the anonymous reviewers, as well as Cleber Jorge Amaral, Jomi Fred Hübner, Esteban Guerrero, Yazan Mualla, Amro Najjar, Helge Spieker, Michael Winikoff, and many others for useful feedback and discussions.
This work was partially supported by the Wallenberg AI, Autonomous Systems and Software Program (WASP) funded by the Knut and Alice Wallenberg Foundation.
%
%
%
\bibliographystyle{splncs04}
\bibliography{references}
\end{document}